\documentclass[a4paper]{article}

\usepackage{graphics} 
\graphicspath{{graphics/}} 
\usepackage{amsmath} 
\usepackage{amssymb}  

\usepackage{mathtools}

\usepackage{amsthm}
\usepackage{microtype}

\usepackage{derivative} 

\newtheorem{thm}{Theorem}
\newtheorem{lemma}[thm]{Lemma}
\newtheorem{rem}[thm]{Remark}

\newtheorem{corollary}[thm]{Corollary}
\newtheorem{definition}[thm]{Definition}

\newcommand{\R}{\mathbb{R}}

\title{\bf
		Stability of Lyapunov redesign trajectory tracking control with unbounded perturbations -- \\A tube-based stability analysis\thanks{This work was supported by Deutsche Forschungsgemeinschaft (DFG) Grant No. 508065537 }
}

\author{
	Niclas Tietze\thanks{Control Engineering Group,  Technische Universt\"at Ilmenau, P.O.~Box 10 05 65, D-98684 Ilmenau, Germany.} \and
	Kai Wulff$^\dag$\thanks{Corresponding author: \texttt{kai.wulff@tu-ilmenau.de}} \and 
	Johann Reger$^\dag$}

\begin{document}

\maketitle
\thispagestyle{empty}
\pagestyle{empty}

\begin{abstract}
Considering a nonlinear system in Byrnes-Isidori form that is subject to
unbounded perturbations, we apply Lyapunov redesign via feedback linearisation for trajectory tracking.
Leveraging the ideas of tube-based geometric characterisation of the invariance properties of the closed loop, we generalise the classical stability criterion from the~literature from constant to nonconstant reference trajectories.
The proposed analysis is tailored to the Lyapunov redesign and the tracking problem insofar as we incorporate the reference trajectory and the transient decrease of the tracking error enforced by the controller.
In particular, we exploit that the Lyapunov function of the tracking error satisfies a differential inequality, thereby guaranteeing that the solution of the closed loop remains in a contracting tube along the reference trajectory.
\end{abstract}

\section{Introduction}
Lyapunov redesign, \cite{EsfK1992}, \cite{MahK1993}, which is also known as the min-max method, \cite{Gut1979}, \cite{CorL1981}, is a well-established robust control technique, which is intrinsically linked to sliding mode control~\cite{EstMF2023}.
The idea of the design is to compensate the influence of the perturbation on the time derivative of a Lyapunov function of the nominal system through an additional control component, thereby guaranteeing stability.
As summarised in Chapter~14 of \cite{Kha2002}, it is well-known that tracking is achieved in presence of perturbations that satisfy a given bounding condition globally.
Moreover, for the special cases of stabilisation and set-point tracking, i.e. constant references, stability is also established for unbounded perturbations, i.e. perturbations that satisfy a given bound only locally on a subset of the state space.
The idea is to show that sets that are positively invariant with respect to the nominal dynamics remain positively invariant under perturbation, \cite{Bla1999}.
Notably, however, tracking is not established for nonconstant references in presence of unbounded perturbations.

\textbf{Contribution:}
We establish the stability of continuous Lyapunov redesign trajectory tracking control with unbounded perturbations.
Given a reference trajectory,
we provide a rigorous local stability analysis and an estimate of the set of admissible initial states for which tracking is achieved.
Conceptually, the stability analysis is comparable to the technique of tube-based (also known as pipe-based or funnel-based) characterisation of the invariance properties of the closed loop, which is commonly applied in reachability analysis and motion control, \cite{JulP2009,HwaJK2024,MajT2017}.
In particular, we show that the Lyapunov redesign enforces that the solution of the perturbed closed loop remains in a tube along the reference.
Leveraging that the reference trajectory is known before run-time, the tube, which is spanned by level sets of the Lyapunov function used of the Lyapunov redesign, is computable offline, thereby facilitating a geometric stability criterion.
Moreover, it turns out that the proposed analysis is able to incorporate the transient decrease of the tracking error through a differential inequality satisfied by the Lyapunov function.
That is, we consider a contracting tube along the reference, showing that the proposed stability criterion conceptually extends beyond positively invariant sets. 
For the special case of a constant reference, i.e. set-point control, the stability criterion simplifies to the classical results from the~literature~\cite{Kha2002}.

\textbf{Structure:} 
This paper is organised as follows.
Section~\ref{sec:problem} gives a definition of the system class and the tracking problem.
Section~\ref{sec:lyapunov_redesign} presents our main results, i.e. the local stability of Lyapunov redesign via a tube-based geometric analysis.
We end by illustrating the results in Section~\ref{sec:example}. 

\section{Problem Definition}\label{sec:problem}
Consider the nonlinear system in Byrnes-Isidori form, \cite{Isi1995},%
\begin{subequations}\label{eq:system}
	\begin{align}
		\dot{\xi} &= A \, \xi + B \big(
		a(\xi,\eta) 
		+ b(\xi,\eta) \, u
		+ \Delta(\xi,\eta,t)
		\big), 
		\label{eq:system_external}
		\\
		\dot{\eta} &= q(\xi,\eta),
		\label{eq:system_internal}
		\\
		y &= \xi_1,
		\label{eq:system_output}
	\end{align}
\end{subequations}
with the external state $\xi(t) = [\xi_1(t),...,\xi_{n_\xi}(t)]^\top \in \mathbb{R}^{n_\xi}$, $\xi(0) =\xi_0$, the internal state $\eta(t) \in \mathbb{R}^{n_\eta}$, $\eta(0) = \eta_0$, and the input $u(t) \in \mathbb{R}$, where $n = n_\xi + n_\eta>0$ for $n_\eta \geq 0$.
The pair $(A,B)$ is in Brunovsk\'{y}-form, i.e. \eqref{eq:system_external} is an integrator chain and relative degree with respect to the output $y$ is $n_\xi \geq 1$.

The internal dynamics \eqref{eq:system_internal} are assumed to be input-to-state stable (ISS) with respect to the input $\xi$, with locally  Lipschitz right-hand side $q:\mathbb{R}^{n_\xi} \times \mathbb{R}^{n_\eta} \mapsto \R^{n_\eta}$.
As shown in \cite{SonW1995}, this is equivalent to the existence of an ISS-Lyapunov function~$V_\eta$ such that, for all $(\xi,\eta) \in \mathbb{R}^{n_\xi} \times \mathbb{R}^{n_\eta}$, 
\begin{subequations}\label{eq:Lyapunov_function_ISS}
	\begin{align}
		&\alpha_1(\Vert \eta \Vert_2) \leq V_\eta(\eta) \leq \alpha_2(\Vert \eta \Vert_2), \label{eq:Lyapunov_function_ISS_norm}
		\\[-0.5ex]
		\pdv{V_\eta}{\eta} \, q(\xi,\eta) &\leq - \alpha_3(\Vert \eta \Vert_2)
		\ \text{ for all } \ \Vert \eta \Vert_2 \geq \gamma(\Vert \xi \Vert_2)
		\label{eq:Lyapunov_function_ISS_norm_derivative}
	\end{align}
\end{subequations}
with class $\mathcal{K}_\infty$ functions $\alpha_1$, $\alpha_2$ and class $\mathcal{K}$ functions $\alpha_3, \, \gamma$.

The known functions $a, \, b: \mathbb{R}^{n_\xi} \times \mathbb{R}^{n_\eta} \mapsto \mathbb{R}$ are continuous, where $|b(\xi,\eta)| \geq  b_0>0$ for all $(\xi,\eta) \in \mathbb{R}^{n_\xi} \times \mathbb{R}^{n_\eta}$. 
The perturbation $\Delta: \mathbb{R}^{n_\xi} \times \mathbb{R}^{n_\eta} \times [0,\infty) \mapsto \mathbb{R}$ is  piecewise continuous with respect to time and locally Lipschitz in $\xi$ and $\eta$.
In~particular, there exist $r>0$ and $\delta \geq 0$ such~that 
\begin{align}
	|\Delta(\xi,\eta,t)| \leq \delta
	\ \, \text{ for all } 
	(\xi,\eta,t) \in \mathcal{D}_r \times\mathcal{P}_r \times [0,\infty),
	\label{eq:pertrubation_bound}
\end{align}
where $\mathcal{D}_r \! \subset \! \mathbb{R}^{n_\xi}$ is an open subset of the ball with radius $r$ and
\begin{align}
	\mathcal{P}_r \coloneqq \Big\{
	\eta \in \R^{n_\eta} \, \big| \,  V_\eta(\eta) < c_r
	\Big\},
	\quad
	c_r \geq \alpha_2\big(\gamma(r)\big).
	\label{eq:invariant_set_internal_dynamics}
\end{align} 

The $n_\xi$ times continuously differentiable reference $y_\mathrm{d}$ and its time derivatives $\dot{y}_\mathrm{d}, \, \ddot{y}_\mathrm{d},...,y_\mathrm{d}^{(n_\xi)}$ are bounded such that the desired state 
\begin{align}\label{eq:desired_state}
	\xi_{\mathrm{d}}(t) \coloneqq  \begin{bmatrix}
		y_\mathrm{d}(t) & \dot{y}_\mathrm{d}(t) & \ddot{y}_\mathrm{d}(t) & ... & y_\mathrm{d}^{(n_\xi - 1)}(t)
	\end{bmatrix}^\top
\end{align}
is contained in $\mathcal{D}_r$ for all $t \geq 0$. 
\begin{definition}[Ultimate Boundedness \cite{Kha2002}]\label{def:ultimate_bound}
	The tracking error $\xi - \xi_{\mathrm{d}}$ is ultimately bounded with ultimate bound $r_\infty > 0$ if there exists some $T_\infty \geq 0$ such that
	\begin{align*}
		\Vert \xi(t) - \xi_{\mathrm{d}}(t) \Vert_2 \leq r_\infty
		\quad \text{ for all } \ t \geq T_\infty.
	\end{align*}
\end{definition}

The goal is to devise a continuous control via Lyapunov redesign \cite{Kha2002} such that the output \eqref{eq:system_output} practically tracks $y_\mathrm{d}$.
That is, the controller shall enforce that the solution $(\xi,\eta)$ is bounded and the tracking error $\bar{\xi} \coloneqq \xi - \xi_{\mathrm{d}}$ is ultimately bounded with an arbitrarily small ultimate bound $r_\infty$, which can be considered as the tracking precision.
Moreover, we shall provide a set of admissible initial states~$(\xi_0,\eta_0)$ for which tracking is~achieved.

\begin{rem}
	In case $n_\xi = n$, the internal dynamics \eqref{eq:system_internal} are dropped.
	In this case, the functions $a, \, b$, and $\Delta$ only depend on $\xi$ and $(\xi,t)$, respectively. 
\end{rem}
\begin{rem}
	Even though we introduce \eqref{eq:system} globally to facilitate the problem definition, our results can be obtained for a local problem setup where the dynamics are satisfied only on the domain $\mathcal{D}_r \times \mathcal{P}_r \subset\R^{n_\xi} \times \R^{n_\eta}$ of interest.
\end{rem}

\section{Main Results}\label{sec:lyapunov_redesign}
Considering Lyapunov redesign, we first present the control design and then establish stability of the closed loop.
We conclude by establishing the connection between the proposed analysis and the well-established results on local stability of Lyapunov redesign from \cite{Kha2002}.
\subsection{Control Design and Closed Loop}
Given
$\bar{\xi} = \xi - \xi_{\mathrm{d}}$, we apply the feedback linearisation
\begin{align}\label{eq:u}
	u = b^{-1}(\xi,\eta)\Big(
	-a(\xi,\eta) + y_\mathrm{d}^{(n_\xi)} + v_\mathrm{N}(\bar{\xi}) + v_\mathrm{L}(\bar{\xi})
	\Big),
\end{align}
where the locally Lipschitz nominal feedback $v_\mathrm{N}$ is chosen such that the dynamics $\dot{\bar{\xi}} = A \, \bar{\xi} + B \, v_\mathrm{N}(\bar{\xi})$
are asymptotically stable with the continuously differentiable, positive definite Lyapunov function $V_\mathrm{N}$ that, for all $\bar{\xi} \in \mathbb{R}^{n_\xi}$, satisfies 
\begin{subequations}\label{eq:redesign_bounds}
	\begin{gather}
		\bar{\alpha}_1(\Vert \bar{\xi} \Vert_2) \leq V_\mathrm{N}(\bar{\xi}) \leq \bar{\alpha}_2(\Vert \bar{\xi} \Vert_2),
		\label{eq:redesign_bounds_norm}
		\\
		\pdv{V_\mathrm{N}}{\bar{\xi}}\big(
		A \, \bar{\xi} + B \, v_\mathrm{N}(\bar{\xi})
		\big) \leq \bar{\alpha}_3(\Vert \bar{\xi} \Vert_2)
		\label{eq:redesign_bounds_derivative}
	\end{gather}
\end{subequations}
with class $\mathcal{K}_\infty$ functions $\bar{\alpha}_1, \, \bar{\alpha}_2$ and class $\mathcal{K}$ function $\bar{\alpha}_3$.
Moreover, the continuous Lyapunov redesign component is
\begin{align}\label{eq:u_redesign}
	v_\mathrm{L}(\bar{\xi}) = - \rho \, 
	\mathrm{sat}\big(\rho \, \mu^{-1} w(\bar{\xi}) \big),
	\quad \
	w(\bar{\xi}) = \pdv{V_\mathrm{N}}{\bar{\xi}} \, B,
\end{align}
with the gain $\rho \geq \delta$ chosen to dominate $\Delta$ from \eqref{eq:pertrubation_bound} and the saturation parameter $\mu > 0$, where we retrieve the typical discontinuous design $v_\mathrm{L}(\bar{\xi}) = - \rho \, \mathrm{sgn}(w(\bar{\xi}))$ for $\mu \rightarrow 0$.

The dynamics of the closed loop \eqref{eq:system}, \eqref{eq:u} read
\begin{subequations}\label{eq:closed_loop}
	\begin{align}
		\dot{\xi} &\!=\! A  \xi \! + \! B \big(
		y_\mathrm{d}^{(n_\xi)}\! \! + \! v_\mathrm{N}(\xi \!-\! \xi_{\mathrm{d}}) \!+\! v_\mathrm{L}(\xi \!- \!\xi_{\mathrm{d}}) \!\!+\! \! \Delta(\xi,\eta,t)
		\big), \!
		\\
		\dot{\eta} &\!=\! q(\xi,\eta).
	\end{align}
\end{subequations}
Moreover, since $\dot{\xi}_\mathrm{d} = A \, \xi_{\mathrm{d}} + B \, y_\mathrm{d}^{(n_\xi)}$, the dynamics of $\bar{\xi}$ are 
\begin{align}\label{eq:closed_loop_error_dynamics}
	\dot{\bar{\xi}} = A \, \bar{\xi} + B \, \big(
	v_\mathrm{N}(\bar{\xi}) + v_\mathrm{L}(\bar{\xi}) + \Delta(\xi_{\mathrm{d}} + \bar{\xi},\eta,t)
	\big).
\end{align}
Following \cite{Kha2002}, it can be verified that the time derivative
\begin{gather*}
	\dot{V}_\mathrm{N} \! = \! \pdv{V_\mathrm{N}}{\bar{\xi}}\big(
	A  \bar{\xi}  \! + \! B v_\mathrm{N}(\bar{\xi})
	\big) \! + \! \pdv{V_\mathrm{N}}{\bar{\xi}}  B\big(
	v_\mathrm{L}(\bar{\xi}) \! + \! \Delta(\xi_{\mathrm{d}} \! + \!  \bar{\xi},\eta,t)
	\big),
\end{gather*}
of $V(\bar{\xi})$, by design in \eqref{eq:redesign_bounds_derivative} and \eqref{eq:u_redesign}, satisfies
\begin{gather}\label{eq:lyapunv_function_extern_derivative}
	\dot{V}_\mathrm{N}(t) \! \leq \! -\bar{\alpha}_3(\Vert \bar{\xi}(t) \Vert_2) \! + \! \tfrac{\mu}{4}
	\quad \text{if} \quad 
	|\Delta(\xi(t),\eta(t),t)| \! \leq  \! \delta.
\end{gather}
Furthermore, the time derivative $\dot{V}_\eta = \pdv{V_\eta}{\eta} \, q(\xi,\eta)$, 
by assumption in \eqref{eq:Lyapunov_function_ISS_norm_derivative}, satisfies
\begin{align}\label{eq:lyapunov_function_intern_derivative}
	\dot{V}_\eta(t) \leq - \alpha_3(\Vert \eta(t) \Vert_2) 
	\quad \text{if} \quad
	\Vert \eta(t) \Vert_2 \geq \gamma(\Vert \xi(t)\Vert_2).
\end{align}

\subsection{Tracking via Tube-Based Analysis}
Note that the estimate \eqref{eq:lyapunv_function_extern_derivative}, together with the bound~\eqref{eq:pertrubation_bound} of~$\Delta$ and the second inequality from \eqref{eq:redesign_bounds_norm}, guarantees that
\begin{align}\label{eq:lyapunv_function_extern_derivative_differential_inequality}
	\dot{V}_\mathrm{N}(t) \leq - \bar{\alpha}_3\Big(
	\bar{\alpha}_2^{-1}\big(
	V_\mathrm{N}(\bar{\xi}(t))
	\big)
	\Big) + \frac{\mu}{4}
\end{align}
if $(\xi(t),\eta(t)) \in \mathcal{D}_r \times \mathcal{P}_r$.
In light of this, let $\nu_{\mu,c_0}(t) \in \mathbb{R}$ with $c_0 \geq 0$ denote the unique solution of the differential~equation
\begin{align}\label{eq:differential_equality_bound_lyapunov_function}
	\dot{\nu}_{\mu,c_0} = - \bar{\alpha}_3\big(\bar{\alpha}_2^{-1}(\nu_{\mu,c_0})\big) + \tfrac{\mu}{4},
	\qquad 
	\nu_{\mu,c_0}(0) = c_0.
\end{align}
Given the desired state $\xi_{\mathrm{d}}$, consider the map $\mathcal{V}_{\xi_{\mathrm{d}},\mu,c_0}$ from the time interval $[0,\infty)$ to the power set (i.e. the set of subsets) of~$\mathbb{R}^{n_\xi}$ that maps $t$ to the compact neighbourhood 
\begin{align}\label{eq:invariant_set_extern}
	\mathcal{V}_{\xi_{\mathrm{d}},\mu,c_0}(t)  \coloneqq  \Big\{
	\xi \in \mathbb{R}^{n_\xi}
	\, \big| \,
	V_\mathrm{N}\big(\xi -  \xi_{\mathrm{d}}(t)\big) \leq \nu_{\mu,c_0}(t)
	\Big\}
\end{align}
of $\xi_{\mathrm{d}}(t)$, which is obtained by centring the level set of $V_\mathrm{N}$ with the level $\nu_{\mu,c_0}(t)$ at $\xi_{\mathrm{d}}(t)$.

Our main result establishes stability whenever $\mu$ and $c_0$ are chosen such that $\mathcal{V}_{\xi_{\mathrm{d}},\mu,c_0}(t)$ is contained in the set $\mathcal{D}_{r}$ of interest for all $t \geq 0$, i.e.
\begin{align}\label{eq:union_in_D}
	\bigcup_{t \geq 0} \mathcal{V}_{\xi_{\mathrm{d}},\mu,c_0}(t) \subseteq \mathcal{D}_r.
\end{align}
Specifically, we establish
tracking and an estimate of the set of admissible initial states by showing that
\begin{align}\label{eq:bound_solution}
	\big(\xi(t),\eta(t)\big) \in \mathcal{V}_{\xi_{\mathrm{d}},\mu,c_0}(t) \times \mathcal{P}_r
	\quad \text{ for all } \, t \geq 0.
\end{align}
\begin{thm}\label{thm:dynamic_analysis}
	Consider the closed loop \eqref{eq:closed_loop} for the trajectory~$\xi_{\mathrm{d}}$ from~\eqref{eq:desired_state}.
	Let $\mu > 0$ and~$c_0 \geq 0$ such that \eqref{eq:union_in_D} is satisfied.
	Then, the solution $(\xi,\eta)$ satisfies~\eqref{eq:bound_solution} for all initial states $(\xi_0,\eta_0) \in \mathcal{V}_{\xi_{\mathrm{d}},\mu,c_0}(0) \times \mathcal{P}_r$.
\end{thm}
\begin{proof}
	Let $[0,T)$ with $T > 0$ (possibly infinity) denote the maximal time interval for which the continuous solution, whose initial state $(\xi_0,\eta_0) \in \mathcal{V}_{\xi_{\mathrm{d}},\mu,c_0}(0) \times \mathcal{P}_r \subseteq \mathcal{D}_r \times \mathcal{P}_r$, is contained in $\mathcal{D}_r \times \mathcal{P}_r$.
	That is, $[0,\infty)$ if $(\xi,\eta)$ remains in $\mathcal{D}_r \times \mathcal{P}_r$, or $[0,T)$ with finite $T$ if $(\xi,\eta)$ approaches the boundary of $\mathcal{D}_r \times \mathcal{P}_r$ as $t \rightarrow T$.  
	Noting that \eqref{eq:lyapunv_function_extern_derivative_differential_inequality} is satisfied for all $t < T$ by construction, we first establish i) $\xi(t) \in \mathcal{V}_{\xi_{\mathrm{d}},\mu,c_0}(t)$ for all $t < T$, and then show that ii) $T$ cannot be finite, thereby yielding \eqref{eq:bound_solution}. 
	
	i) The differential inequality \eqref{eq:lyapunv_function_extern_derivative_differential_inequality}, which corresponds to the differential equation \eqref{eq:differential_equality_bound_lyapunov_function}, is satisfied for all $t < T$.
	Moreover, $V_\mathrm{N}(\bar{\xi}(0)) \leq  \nu_{\mu,c_0}(0) = c_0$ since $\xi_0 \in \mathcal{V}_{\xi_{\mathrm{d}},\mu,c_0}(0)$ by assumption.
	Thus, applying (the comparison) Lemma~\ref{lem:comparison_lemma} from the Appendix, we obtain $V_\mathrm{N}(\bar{\xi}(t)) \leq \nu_{\mu,c_0}(t)$ for all $t < T$.
	Consequently, $\xi(t) \in \mathcal{V}_{\xi_{\mathrm{d}},\mu,c_0}(t)$ by construction~\eqref{eq:invariant_set_extern}.
	
	ii) For proof by contradiction, suppose that $T$ is finite.
	Then, by definition, at least one of the following conditions is satisfied as $t \rightarrow T$: a) $\xi(t)$ gets arbitrarily close to the boundary of $\mathcal{D}_r$, b) $\eta(t)$ gets arbitrarily close to the boundary of $\mathcal{P}_r$. 
	To exclude a) and b), note that $\mathcal{V}_{\xi_{\mathrm{d}},\mu,c_0}(t) \subseteq \bigcup_{t \geq 0} \mathcal{V}_{\xi_{\mathrm{d}},\mu,c_0}(t)$ for all $t \geq 0$ by construction, where $\mathcal{V}_{\xi_{\mathrm{d}},\mu,c_0} \subseteq \mathcal{D}_r$ by assumption in \eqref{eq:union_in_D}.
	Let $d_\mathrm{m}(t)$ denote the minimal distance between the boundary of $\mathcal{V}_{\xi_{\mathrm{d}},\mu,c_0}(t) \subseteq \mathcal{D}_r$ and $\mathcal{D}_r$.
	Applying the extreme value theorem to $d_\mathrm{m}$, which is continuous since~$\xi_{\mathrm{d}}$ and $\nu_{\mu,c_0}$ are continuous, there exists $t_\mathrm{m} \in [0,T]$ such that $d_\mathrm{m}(t) \geq d_\mathrm{m}(t_\mathrm{m})$ for all $t \in [0,T]$.
	However, the compact set $\mathcal{V}_{\xi_{\mathrm{d}},\mu,c_0}(t_\mathrm{m})$ is contained in the open set $\mathcal{D}_r$ by assumption.
	Thus, applying the Lebesgue number lemma (see e.g. \cite{Roy2015}), $d_\mathrm{m}(t_\mathrm{m}) > 0$.
	Consequently, $d_\mathrm{m}$ is bounded away from zero in $[0,T]$, thereby showing that $\xi(t) \in \mathcal{V}_{\xi_{\mathrm{d}},\mu,c_0}(t) \subset \mathcal{D}_r$ cannot get arbitrarily close to the boundary of $\mathcal{D}_r$ as $t \rightarrow T$ for finite $T$, which excludes~a). 
	In particular, since $\mathcal{D}_r$ is contained in the ball with radius $r$ by assumption, there exists some positive $r_\mathrm{m} < r$ such that $\Vert \xi(t) \Vert \leq r_\mathrm{m}$ for all $t < T$.
	Moreover, the time derivative of $V_\eta(\eta)$ satisfies~\eqref{eq:lyapunov_function_intern_derivative}.
	Thus, together with~\eqref{eq:Lyapunov_function_ISS_norm} and the class $\mathcal{K}$ function $\gamma$, we obtain $\dot{V}_\eta(t) \leq 0$ whenever $V_\eta(\eta(t)) \geq \alpha_2(\gamma(r_\mathrm{m}))$.
	Consequently, $V_\eta(t) \leq \max\{
	V_\eta(\eta_0), \alpha_2(\gamma(r_\mathrm{m}))
	\}$ for all $t < T$.
	However, $V_\eta(\eta_0) < c_r$ by assumption $\eta_0 \in \mathcal{P}_r$ and $\alpha_2(\gamma(r_\mathrm{m})) <  \alpha_2(\gamma(r)) \leq c_r$ by construction of $c_r$ in~\eqref{eq:invariant_set_internal_dynamics} with class $\mathcal{K}$ functions $\alpha_2$ and~$\gamma$.
	Therefore, $V_\eta(\eta)$ is bounded away from $c_r$, i.e. $\eta$ cannot get arbitrarily close to the boundary of
	\eqref{eq:invariant_set_internal_dynamics} in finite time $T$, thereby excluding~b).

	With contradiction in a) and b), we obtain $T = \infty$.
	Finally, together with $\xi(t) \in \mathcal{V}_{\xi_{\mathrm{d}},\mu,c_0}(t)$,	\eqref{eq:bound_solution} is satisfied.
\end{proof}
Theorem~\ref{thm:dynamic_analysis} establishes that the Lyapunov function $V_\mathrm{N}(\xi - \xi_{\mathrm{d}})$, whose time derivative satisfies the differential inequality~\eqref{eq:lyapunv_function_extern_derivative_differential_inequality} by design, decreases faster than the solution $\nu_{\mu,c_0}$ of the corresponding differential equation~\eqref{eq:differential_equality_bound_lyapunov_function}.
Specifically, we apply the comparison Lemma to incorporate the decrease of the tracking error into the analysis, thereby yielding the bound $	\mathcal{V}_{\xi_{\mathrm{d}},\mu,c_0}(t) \ni \xi(t)$ for all $t \geq 0$.
Moreover, given that $\xi$ remains within the union \eqref{eq:union_in_D}, which is contained in the ball with radius~$r$ by assumption, $\mathcal{P}_r$ is positively invariant with respect to the ISS internal dynamics~\eqref{eq:system_internal}.

Given the resulting
bound \eqref{eq:bound_solution},
two aspects are of particular interest.
On the one hand, due to the first inequality of \eqref{eq:redesign_bounds_norm}, the tracking error satisfies
\begin{align*}
	\Vert \bar{\xi}(t) \Vert_2 
	\leq \bar{\alpha}_1^{-1}\big(
	V_\mathrm{N}(\bar{\xi}(t))
	\big) 
	\leq \bar{\alpha}_1^{-1}\big(
	\nu_{\mu,c_0}(t)
	\big)
	\, \text{ for all } \, t \geq 0,
\end{align*}
where the solution
of \eqref{eq:differential_equality_bound_lyapunov_function} satisfies \begin{align}\label{eq:limit_bound}
	\textstyle
	\lim_{t \rightarrow \infty} \nu_{\mu,c_0}(t) = \bar{\alpha}_2\big(\bar{\alpha}_3^{-1}(\tfrac{\mu}{4})\big) \eqqcolon \bar{\alpha}_\infty(\mu)
\end{align}
with the class $\mathcal{K}$ function $\bar{\alpha}_\infty$.
Thus, \eqref{eq:bound_solution} guarantees practical tracking.
In particular, recalling Definition~\ref{def:ultimate_bound}, it is readily verified that choosing 
\begin{align}\label{eq:mu_desired_precision}
	\mu < \bar{\alpha}_\infty^{-1}\big(
	\bar{\alpha}_1(r_\infty)
	\big)
\end{align}
guarantees that $\bar{\xi}$ is ultimately bounded with the desired ultimate bound $r_\infty > 0$.
Notably, we can enforce arbitrarily good precision by decreasing the parameter $\mu \rightarrow 0$. 
On the other hand, the set $\mathcal{V}_{\xi_{\mathrm{d}},\mu,c_0}(0) \times \mathcal{P}_r$ is an estimate of the set of admissible initial states for which tracking is achieved.

\begin{rem}
	Given that~\eqref{eq:bound_solution} is satisfied for all $(\xi_0,\eta_0) \in \mathcal{V}_{\xi_{\mathrm{d}},\mu,c_0}(0) \times \mathcal{P}_r$, the set $\mathcal{V}_{\xi_{\mathrm{d}},\mu,c_0}(t)$ is an estimate of the set of states the external dynamics may evolve to in time $t$, given $\xi_0 \in \mathcal{V}_{\xi_{\mathrm{d}},\mu,c_0}(0)$.
	In the context of reachability analysis and motion control, this set is known as the reach set \cite{YazP2004}.
	In light of this geometric interpretation (see Section~\ref{sec:example}), the map $\mathcal{V}_{\xi_{\mathrm{d}},\mu,c_0}$ corresponds to a tube along the reference trajectory~$\xi_{\mathrm{d}}$,~\cite{MajT2017}.
	Geometrically, \eqref{eq:union_in_D} guarantees that the tube, which is computed before run-time, is contained in $\mathcal{D}_r$.
	In other words, Theorem~\ref{thm:dynamic_analysis} preservers the fundamental idea of the classical Lyapunov-based stability analysis, namely to guarantee the boundedness of the solution~$(\xi,\eta)$ of the closed loop by checking a geometric condition that can be evaluated before run-time, i.e. without solving the state equation containing the perturbation~$\Delta$.    
	Specifically, the differential equation \eqref{eq:lyapunv_function_extern_derivative_differential_inequality} does not contain $\Delta$. 
\end{rem}

Computing $\mathcal{V}_{\xi_{\mathrm{d}},\mu,c_0}$ requires solving the differential equation \eqref{eq:differential_equality_bound_lyapunov_function}.
However, noting that the solution $\nu_{\mu,c_0}$ of \eqref{eq:differential_equality_bound_lyapunov_function} is nonincreasing for all $c_0 \geq \bar{\alpha}_\infty(\mu)$ by design,
an overestimate of the bound $\mathcal{V}_{\xi_{\mathrm{d}},\mu,c_0}(t) \ni \xi(t)$ is given by
\begin{align}\label{eq:invariant_set_extern_static}
	\mathcal{W}_{\xi_{\mathrm{d}},c_0}(t) \coloneqq  \Big\{
	\xi \in \mathbb{R}^{n_\xi}
	\, \big| \,
	V_\mathrm{N}\big(\xi -  \xi_{\mathrm{d}}(t)\big) \leq c_0 
	\Big\}, 
\end{align}
where $\mathcal{W}_{\xi_{\mathrm{d}},c_0}(0) = \mathcal{V}_{\xi_{\mathrm{d}},\mu,c_0}(0)$.
That is,
\begin{align*}
	\mathcal{W}_{\xi_{\mathrm{d}},c_0}(t) \supseteq \mathcal{V}_{\xi_{\mathrm{d}},\mu,c_0}(0)
	\quad \text{ for all } \ t \geq 0. 
\end{align*}
Thus, a sufficient condition for \eqref{eq:union_in_D} is
\begin{align}\label{eq:union_in_D_static}
	\bigcup_{t \geq 0} \mathcal{W}_{\xi_{\mathrm{d}},c_0}(t) \subseteq \mathcal{D}_r.
\end{align}
\begin{corollary}\label{cor:static_analysis}
	Consider the closed loop \eqref{eq:closed_loop} for the trajectory~$\xi_{\mathrm{d}}$ from~\eqref{eq:desired_state}.
	Let $\mu > 0$ and $c_0 \geq \bar{\alpha}_\infty(\mu)$ such that~\eqref{eq:union_in_D_static} is satisfied.
	Then, the solution $(\xi,\eta)$ satisfies \eqref{eq:bound_solution}  for all $(\xi_0,\eta_0) \in \mathcal{W}_{\xi_{\mathrm{d}},c_0}(0) \times \mathcal{P}_r$.
\end{corollary}
\begin{proof}
	We have $(\xi_0,\eta_0) \in \mathcal{W}_{\xi_{\mathrm{d}},c_0}(0) \times \mathcal{P}_r = \mathcal{V}_{\xi_{\mathrm{d}},\mu,c_0}(0) \times \mathcal{P}_r$ and $\bigcup_{t \geq 0} \mathcal{V}_{\xi_{\mathrm{d}},\mu,c_0}(t) \subseteq \bigcup_{t \geq 0} \mathcal{W}_{\xi_{\mathrm{d}},c_0}(t) \subseteq \mathcal{D}_r$.
	Apply Theorem~\ref{thm:dynamic_analysis}.
\end{proof}
Similar to Theorem~\ref{thm:dynamic_analysis}, Corollary~\ref{cor:static_analysis} establishes tracking via a geometric condition.
In particular, we require the tube given by the map $	\mathcal{W}_{\xi_{\mathrm{d}},c_0}$ to be contained in $\mathcal{D}_r$.
However, in contrast to Theorem~\ref{thm:dynamic_analysis}, the differential equation \eqref{eq:differential_equality_bound_lyapunov_function} need not be solved to evaluate Corollary~\ref{cor:static_analysis}, thereby facilitating the evaluation.
The idea is to overestimate $\mathcal{V}_{\xi_{\mathrm{d}},\mu,c_0}$ with $\mathcal{W}_{\xi_{\mathrm{d}},c_0}$ by overestimating  $\nu_{\mu,c_0}(t)$ with $\nu_{\mu,c_0}(0) = c_0$.
In other words, we avoid solving \eqref{eq:differential_equality_bound_lyapunov_function} at the expense of a more conservative result. 
Specifically, Corollary~\ref{cor:static_analysis} is more conservative  than Theorem~\ref{thm:dynamic_analysis} insofar as we require $c_0 \geq \bar{\alpha}_\infty(\mu)$ to exclude that~$\nu_{\mu,c_0}$ increases and \eqref{eq:union_in_D_static} is more restrictive than \eqref{eq:union_in_D}.
Note that, even tough \eqref{eq:invariant_set_extern_static} does not explicitly depend on $\mu$, the requirement $c_0 \geq \bar{\alpha}_\infty(\mu)$ induces an implicit dependence on $\mu$.
\begin{rem}[Finite-Time Analysis]
	Theorem~\ref{thm:dynamic_analysis} establishes ultimate boundedness considering the union \eqref{eq:union_in_D} of $\mathcal{V}_{\xi_{\mathrm{d}},\mu,c_0}(t)$ over the entire time horizon $[0,\infty)$.
	However, following the idea of (finite-time) reachability analysis, the analysis can readily be adapted to a finite time interval $[0,T_\mathrm{f}]$, $T_\mathrm{f} > 0$.
	In particular, by minor adaptions of the proof of Theorem~\ref{thm:dynamic_analysis}, it can be shown that the solution $(\xi,\eta)$ satisfies
	\begin{align*}
		\big(\xi(t),\eta(t)\big) \in \mathcal{W}_{\xi_{\mathrm{d}},c_0}(t) \times \mathcal{P}_r \quad \text{ for all } \, t \in [0,T_\mathrm{f}]
	\end{align*}
	for every $(\xi_0,\eta_0) \in \mathcal{V}_{\xi_{\mathrm{d}},\mu,c_0}(0) \times \mathcal{P}_r$ whenever
	\begin{align*}
		\bigcup_{t \in [0,T_\mathrm{f}]} \mathcal{V}_{\xi_{\mathrm{d}},\mu,c_0}(t) \subseteq \mathcal{D}_r.
	\end{align*}
	That is, for boundedness in $[0,T_\mathrm{f}] \ni t$, we require only the part of the tube that corresponds to $t \leq T_\mathrm{f}$ to be contained in $\mathcal{D}_r$.
	This adaption is less conservative than Theorem~\ref{thm:dynamic_analysis}, as it covers the case where $(\xi,\eta)$ does not remain in $\mathcal{D}_r \times \mathcal{P}_r$ for all $t \geq 0$ i.e. only for a finite time. 
	In particular, the finite-time analysis facilitates a lower bound for the time at which the solution leaves the set of interest.
		Notably, this interpretation facilitates the connection to motion control \cite{JulP2009,HwaJK2024,MajT2017}, where the idea is to apply the similar tube-based finite-time stability analysis for optimisation-based control in the context of receding horizon trajectory paling.
\end{rem}
\begin{rem}[Discontinuous Case]
	We obtain the conventional discontinuous design $v_\mathrm{L}(\bar{\xi}) = - \rho \, \mathrm{sgn}(w(\bar{\xi}))$ in the limit $\mu \rightarrow 0$. 
	In this case, choosing $c_0 = 0$ yields $\nu_{\mu,c_0} \equiv 0$ and
	\begin{align*}
		\mathcal{V}_{\xi_{\mathrm{d}},\mu,c_0}(t) 
		= \mathcal{W}_{\xi_{\mathrm{d}},c_0}(t) 
		= \big\{\xi_{\mathrm{d}}(t)\big\}
	\end{align*} 
	is the set that contains the single element $\xi_{\mathrm{d}}(t)$.
	Thus, evaluating Theorem~\ref{thm:dynamic_analysis}, we obtain $\xi(t) = \xi_{\mathrm{d}}(t)$ for all~$t \geq 0$ whenever $\xi_0 = \xi_{\mathrm{d}}(0)$ since $\bigcup_{t \geq 0}\{\xi_{\mathrm{d}}(t)\} \subseteq \mathcal{D}_r$ by assumption on $\xi_{\mathrm{d}}$ from \eqref{eq:desired_state}.
	In other words, the analysis captures that the design enforces exact tracking $\xi \equiv \xi_{\mathrm{d}}$ if the initial value $\xi_0 = \xi_{\mathrm{d}}(0)$ of the process lies on the reference trajectory, i.e. the insensitivity of the discontinuous Lyapunov redesign with respect to $\Delta$.

\end{rem}
\begin{rem}[Minkowski Sum]
	Inspired by \cite{TieWR2024CDCSMC} and \cite{HwaJK2024}, the geometric interpretation of the analysis is further emphasised by the Minkowski sum \cite{BerCK2008}, which is a set operation e.g. used in model predictive control~\cite{KouC2015}. 
	In particular, introducing 
	\begin{align*}
		\bar{\mathcal{V}}_{\mu,c_0}(t) \coloneqq \Big\{
		\bar{\xi} \in \mathbb{R}^{n_\xi}
		\, \big| \, 
		V_\mathrm{N}(\bar{\xi}) \leq \nu_{\mu,c_0}(t)
		\Big\},
	\end{align*}
	the set $\mathcal{V}_{\xi_{\mathrm{d}},\mu,c_0}(t)  =  \{
	\xi_\mathrm{d}(t)
	\} \oplus \bar{\mathcal{V}}_{\mu,c_0}(t)$.
	In contrast to \eqref{eq:invariant_set_extern}, the sum representation does not explicitly introduce $\mathcal{V}_{\xi_{\mathrm{d}},\mu,c_0}(t)$ as a subset of the state space $\mathbb{R}^{n_\xi} \ni \xi(t)$ of the process.
	Rather, $\mathcal{V}_{\xi_{\mathrm{d}},\mu,c_0}(t)$ is constructed as the sum of the desired state $\xi_{\mathrm{d}}(t)$ and the set $\bar{\mathcal{V}}_{\mu,c_0}(t)$, which is defined in the state space $\mathbb{R}^{n_\xi} \ni \bar{\xi}(t)$ of the tracking error.
	Analogously, for the static analysis, $\mathcal{W}_{\xi_{\mathrm{d}},c_0}(t) =  \{
	\xi_\mathrm{d}(t)
	\}  \oplus \bar{\mathcal{V}}_{\mu,c_0}(0)$.

	For every $c_0 \geq \bar{\alpha}_\infty(\mu)$, the set  $\bar{\mathcal{V}}_{\mu,c_0}(t) \subseteq \bar{\mathcal{V}}_{\mu,c_0}(0)$ for all $t \geq 0$ since $\nu_{\mu,c_0}$ is nonincreasing.
	That is, $\bar{\mathcal{V}}_{\mu,c_0}(0)$ is positively invariant with respect to 
	\eqref{eq:closed_loop_error_dynamics}. 
	This shows that, Theorem~\ref{thm:dynamic_analysis}, which incorporates $\bar{\mathcal{V}}_{\mu,c_0}(t)$ instead of $\bar{\mathcal{V}}_{\mu,c_0}(0)$, facilitates less conservative results by conceptually extending beyond positively invariant sets.
\end{rem}

\subsection{Special Case: Set-Point Tracking}
Consider the constant desired state
\begin{align}\label{eq:set_point}
	\xi_{\mathrm{d}} = \begin{bmatrix}
		y_\mathrm{d} & 0 & ... & 0
	\end{bmatrix}^\top \in \mathcal{D}_r.
\end{align}
Given that $\mu$ can be chosen arbitrarily small, we assume without loss of generality 
$c_0 \geq \bar{\alpha}_\infty(\mu)$.
Then, 
\begin{align*}
	\mathcal{V}_{\xi_{\mathrm{d}},\mu,c_0}(t) \subseteq \mathcal{V}_{\xi_{\mathrm{d}},\mu,c_0}(0)
	\quad \text{ for all } \, t \geq 0
\end{align*}
since $\nu_{\mu,c_0}$ is nonincreasing.
Therefore, 
\begin{align}\label{eq:set_point_simplified_union}
	\bigcup_{t \geq 0} \mathcal{V}_{\xi_{\mathrm{d}},\mu,c_0}(t) = \mathcal{V}_{\xi_{\mathrm{d}},\mu,c_0}(0),
\end{align}
simplifying Theorem~\ref{thm:dynamic_analysis} as follows.
\begin{corollary}[Set-Point Control]\label{cor:set_point}
	Consider the closed loop~\eqref{eq:closed_loop} for the set-point~$\xi_{\mathrm{d}}$ from~\eqref{eq:set_point}.
	Let $\mu > 0$ and $c_0 \geq \bar{\alpha}_\infty(\mu)$ such that $\mathcal{V}_{\xi_{\mathrm{d}},\mu,c_0}(0) \subseteq \mathcal{D}_r$.
	Then, the solution $(\xi,\eta)$ satisfies~\eqref{eq:bound_solution} for all $(\xi_0,\eta_0) \in \mathcal{V}_{\xi_{\mathrm{d}},\mu,c_0}(0) \times \mathcal{P}_r$.
\end{corollary}
Furthermore, considering the set-point~\eqref{eq:set_point}, the set $\mathcal{W}_{\xi_{\mathrm{d}},c_0}(t) = \mathcal{W}_{\xi_{\mathrm{d}},c_0}(0)$ for all $t \geq 0$.
Thus,
\begin{align}\label{eq:set_point_simplified_bounds}
	\mathcal{V}_{\xi_{\mathrm{d}},\mu,c_0}(0)
	= \mathcal{W}_{\xi_{\mathrm{d}},c_0}(0)
	= \bigcup_{t \geq 0} \mathcal{W}_{\xi_{\mathrm{d}},c_0}(t),
\end{align}
which shows that the conditions under which Theorem~\ref{thm:dynamic_analysis} and  Corollary~\ref{cor:static_analysis} establish tracking coincide.
In other words, explicitly interoperating the solution $\nu_{\mu,c_0}$ of the differential equation \eqref{eq:differential_equality_bound_lyapunov_function} does not yield a less conservative estimate of the region of attraction for constant references.

Corollary~\ref{cor:set_point} establishes practical tracking.
In particular, choosing $\mu$ as in \eqref{eq:mu_desired_precision} enforces the arbitrarily small ultimate bound $r_\infty$ for $\bar{\xi}$.
	This shows that the result is an adaption of the well-established results on the local stability of Lyapunov redesign found in Section 14.2 of \cite{Kha2002} in the sense that we guarantee practical tracking for sufficiently small $\mu$ satisfying~\eqref{eq:mu_desired_precision}.
	In other words, the proposed tube-based stability analysis generalises the well-established Lyapunov-based stability analysis from set-point control to trajectory tracking.

\begin{rem}[Availability of the Reference]
	The controller~\eqref{eq:u}	requires the signals $\dot{y}_\mathrm{d}, \, \ddot{y}_\mathrm{d},...,y_\mathrm{d}^{(n_\xi)}$ to be known only at run-time.
	However, we require $\xi_\mathrm{d}$ to be known before run-time to conduct an a priori analysis of the stability of the closed loop.
	In other words, we require $\xi_{\mathrm{d}}$ to be known before run-time since the stability analysis incorporates $\xi_{\mathrm{d}}$.
	Notably, this is trivially satisfied for the special case of set-point tracking and stabilisation of the origin, where the constant reference trajectory \eqref{eq:set_point} is available before run-time.
	In essence, the availability of~$\xi_{\mathrm{d}}$ before run-time imposes additional restrictions only for nonconstant $y_{\mathrm{d}}$.
\end{rem}

\section{Illustrative Example}\label{sec:example}
Consider system \eqref{eq:system} with $n_\xi = 2$, $n_\eta = 1$ for $a(\xi,\eta) = 0$, $b(\xi,\eta) \! =  1$, and $q(\xi,\eta) = - \eta + \xi_1$, which satisfies \eqref{eq:Lyapunov_function_ISS} with $V_\eta(\eta) = \alpha_1(|\eta|) = \alpha_2(|\eta|) = \frac{1}{2} \, \eta^2$ and $\gamma(\Vert \xi \Vert_2) = \theta_\eta^{-1} \Vert \xi \Vert_2$ for $\theta_\eta \in (0,1)$.
Consider the set
\begin{align*}
	\mathcal{D}_r = \Big\{
	\xi \in \mathbb{R}^2 
	\big|
	|\xi_1| < 1 \wedge |\xi_2| < 1
	\Big\},
\end{align*}
which is contained in the ball with radius $r = \sqrt{2}$.
Then, choosing $c_r = \alpha_2(\gamma(r))$ in \eqref{eq:invariant_set_internal_dynamics}, the set $\mathcal{P}_r = (-\theta_\eta^{-1} r, \, \theta_\eta^{-1}r)$, where we chose $\theta_\eta = 0.95$.
Consider an arbitrary perturbation~$\Delta$ that satisfies \eqref{eq:pertrubation_bound} with $\delta = 1$.
For simulation, we apply the specific perturbation $\Delta(\xi,\eta,t) = \frac{1}{8}(\xi_1^2 + \xi_2^2) + \frac{3}{4} \sin(5 \, t)$.%

Let $\rho = 1$ and $\mu = 10^{-3}$.
Considering the typical linear design, choose the gain $k = \begin{bsmallmatrix}
	1 \\ 2
\end{bsmallmatrix}$ of $v_\mathrm{N}(\bar{\xi}) = -k^\top \bar{\xi}$ so that $\bar{A} = A - B \, k^\top$ is Hurwitz.
Given the solution $P = \frac{1}{2} \begin{bsmallmatrix}
	3 & 1 \\ 1 & 1
\end{bsmallmatrix}$ of 
$\bar{A}^\top P + P \, \bar{A} = -I$, $V_\mathrm{N}(\bar{\xi}) = \bar{\xi}^{\, \top} P \, \bar{\xi}$ satisfies \eqref{eq:redesign_bounds} with $\bar{\alpha}_1(\Vert \bar{\xi} \Vert_2) = \lambda_\mathrm{min}(P) \, \Vert \bar{\xi} \Vert_2^2$, $\bar{\alpha}_2(\Vert \bar{\xi} \Vert_2) = \lambda_\mathrm{max}(P) \, \Vert \bar{\xi} \Vert_2^2$, and $\bar{\alpha}_3(\Vert \bar{\xi} \Vert_2) = \Vert \bar{\xi} \Vert_2^2$.
Thus, the solution of \eqref{eq:differential_equality_bound_lyapunov_function} is
\begin{align*}
	\nu_{\mu,c_0}(t) = \lambda_{\mathrm{max}}(P)\big(
	\tfrac{\mu}{4} - e^{-\lambda_{\mathrm{max}}^{-1}(P) \, t}(
	\tfrac{\mu}{4} - \lambda_{\mathrm{max}}^{-1}(P) \, c_0
	)
	\big).
\end{align*}
Moreover, the sets \eqref{eq:invariant_set_extern} and \eqref{eq:invariant_set_extern_static} are the ellipsoids
\begin{gather*}
	\mathcal{V}_{\xi_{\mathrm{d}},\mu,c_0}(t) \! = \! \Big\{ \!
	\xi  \! \in \mathbb{R}^{2}
	\big|
	\big(
	\xi - \xi_{\mathrm{d}}(t)
	\big)^{\! \top} \! P \big(
	\xi - \xi_{\mathrm{d}}(t)
	\big) \! \leq \nu_{\mu,c_0}(t)
	\! \Big\},
	\\
	\mathcal{W}_{\xi_{\mathrm{d}},c_0}(t) = \Big\{
	\xi \in  \mathbb{R}^{2}
	\, \big| \,
	\big(
	\xi - \xi_{\mathrm{d}}(t)
	\big)^{\! \top} \! P \big(
	\xi - \xi_{\mathrm{d}}(t)
	\big) \! \leq c_0
	\Big\}.
\end{gather*} 
\begin{figure}
	\centering
	\includegraphics[width=.7\linewidth]{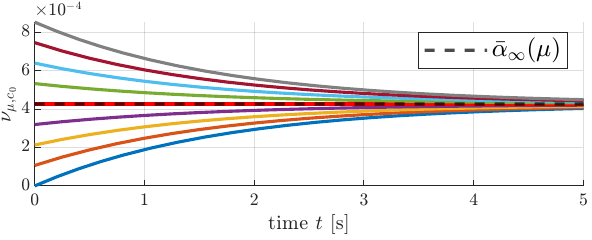}
	\vspace{-2ex}
	\caption{
		Solution $\nu_{\mu,c_0}$ of \eqref{eq:differential_equality_bound_lyapunov_function}
		for increasing~$c_0 = \nu_{\mu,c_0}(0)$.
	}
	\label{fig:lyapunov_redesign_ode_sol}
\end{figure}%
Figure~\ref{fig:lyapunov_redesign_ode_sol} shows $\nu_{\mu,c_0}$ for increasing $c_0$.
Notably, the plot is representative for the solution of \eqref{eq:differential_equality_bound_lyapunov_function} insofar as $\nu_{\mu,c_0}$ is i) strictly monotonically increasing to the limit \eqref{eq:limit_bound} for $c_0 < \bar{\alpha}_\infty(\mu)$, ii) constant for $c_0 = \bar{\alpha}_\infty(\mu)$, and iii) strictly monotonically decreasing for $c_0 > \bar{\alpha}_\infty(\mu)$.

\subsection{Trajectory Tracking}
\begin{figure}
	\centering
	\includegraphics[width=.8\linewidth]{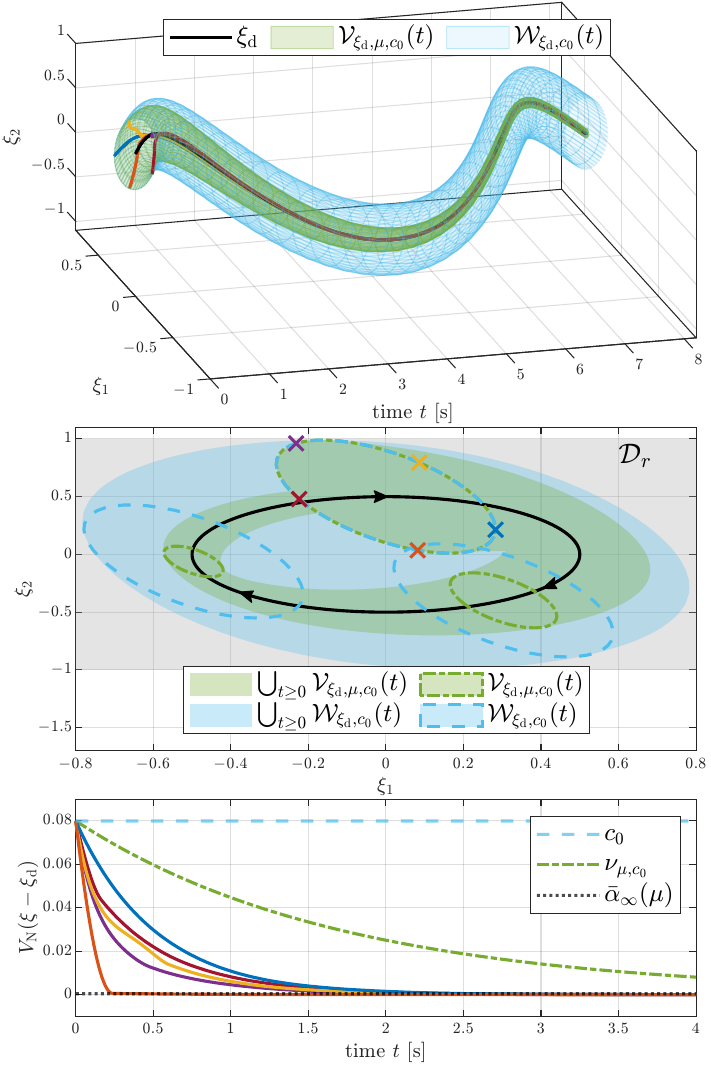}
	\caption{
		Tube-based stability analysis for a sinusoidal reference, shown in the time domain (top and bottom) and the phase plane (middle).
	}
	\label{fig:lyapunov_redesign}
\end{figure}

Let the reference trajectory be given by $y_\mathrm{d}(t) = 0.5 \, \sin(t)$.
The top plot of Figure~\ref{fig:lyapunov_redesign} shows the evolution of the external state $\xi$.
The set $\mathcal{V}_{\xi_{\mathrm{d}},\mu,c_0}(t)$ spans a tube along $\xi_{\mathrm{d}}$, whose size decreases over time, where~$\mu$ determines the size for~$t \rightarrow \infty$.
We obtain the union \eqref{eq:union_in_D} by projecting the tube onto the $\xi_1$-$\xi_2$-plane, as shown in the middle plot.
There, the set $\mathcal{V}_{\xi_{\mathrm{d}},\mu,c_0}(t)$ is shown for the three time instances $t \in \{0,1.5,4.5\} \,$s using dash-dotted lines.
As~$t$ increases, the contracting set moves along $\xi_{\mathrm{d}}$ clockwise, forming the boundary of~\eqref{eq:union_in_D}.
Notably,
$c_0 = 0.08$ is chosen such that \eqref{eq:union_in_D} is satisfied, i.e. the union is a subset of $\mathcal{D}_r$, which is shaded in grey.
Thus, Theorem~\ref{thm:dynamic_analysis} guarantees tracking for all $(\xi_0,\eta_0) \in \mathcal{V}_{\xi_{\mathrm{d}},\mu,c_0}(0) \times \mathcal{P}_r$.
For demonstration, the top plot shows the solution $\xi$ of the closed loop for five different~$\xi_0$ on the boundary of $\mathcal{V}_{\xi_{\mathrm{d}},\mu,c_0}(0)$ (marked by crosses in the middle plot) and $\eta_0 = 1.4 \in \mathcal{P}_r$.
Each solution remains in the contracting tube, i.e. $V_\mathrm{N}(\bar{\xi}) \leq \nu_{\mu,c_0}$ as is shown in the bottom plot.
For comparison,
Figure~\ref{fig:lyapunov_redesign} also shows the non-contracting tube given by $\mathcal{W}_{\xi_{\mathrm{d}},c_0}(t)$.
Notably, $\mathcal{W}_{\xi_{\mathrm{d}},c_0}(t)$ overestimates $\mathcal{V}_{\xi_{\mathrm{d}},\mu,c_0}(t)$ via a tube of constant size along $\xi_{\mathrm{d}}$.
Corollary~\ref{cor:static_analysis} guarantees tracking for the depicted $\xi_0$ since \eqref{eq:union_in_D_static} is satisfied.

\begin{figure}
	\centering
	\includegraphics[width=.8\linewidth]{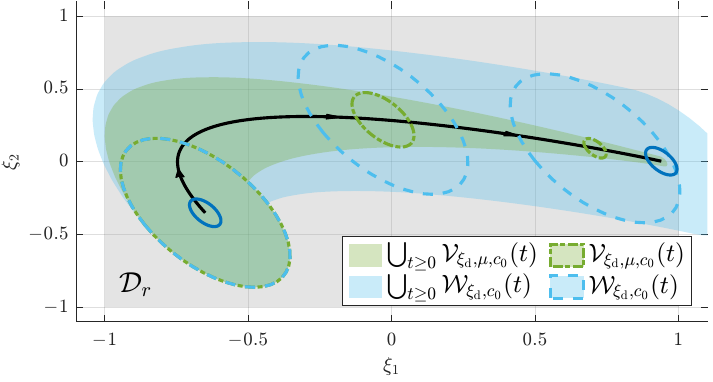}
	\caption{
		Tube-based stability analysis for a transition. 
		Incorporating the transient decrease of the tracking error enforced by the controller reduces the conservatism of the stability criterion.
	}
	\label{fig:lyapunov_redesign_transition}
\end{figure}

To illustrate the reduced conservatism of Theorem~\ref{thm:dynamic_analysis} over Corollary~\ref{cor:static_analysis}, Figure~\ref{fig:lyapunov_redesign_transition} shows the $\xi_1$-$\xi_2$-plane analysis of Figure~\ref{fig:lyapunov_redesign}, but for $y_\mathrm{d}(t) = 0.95 - (1.15 \, t + 1.6) \, e^{- 0.5 \, t}$, which can be considered as a smooth transition from $y_\mathrm{d}(0) = -0.65$ to $\lim_{t \rightarrow \infty} y_\mathrm{d}(t) = 0.95$, which is close to the boundary of the set of interest $\mathcal{D}_r$.
Since~$c_0 = 0.0875$ is chosen to satisfy~\eqref{eq:union_in_D}, Theorem~\ref{thm:dynamic_analysis} guarantees stability.
However, \eqref{eq:union_in_D_static} is not satisfied since the blue tube leaves $\mathcal{D}_r$ both to the left (for small times) and to the right (for $t\rightarrow \infty$).
Conceptually, Theorem~\ref{thm:dynamic_analysis} is less conservative since the incorporation of the transient decrease of $\bar{\xi}$ allows $y_\mathrm{d}$ to be close to the boundary of $\mathcal{D}_r$ for (large) times at which $\bar{\xi}$ decreased sufficiently.
For comparison, the plot further shows $\mathcal{W}_{\xi_{\mathrm{d}},c_0}(t)$ with $c_0 = 0.03$ for $t = 0$ and $t \rightarrow \infty$ using a dark-blue solid line.
Notably, $\mathcal{W}_{\xi_{\mathrm{d}},c_0}(0)$ can be understood as an estimate of the largest set for which Corollary~\ref{cor:static_analysis} guarantees stability since $\mathcal{W}_{\xi_{\mathrm{d}},c_0}(t)$ approaches the boundary of $\mathcal{D}_r$ as $t \rightarrow \infty$. 

\subsection{Set-Point Tracking}
\begin{figure}
	\centering
	\includegraphics[width=.8\linewidth]{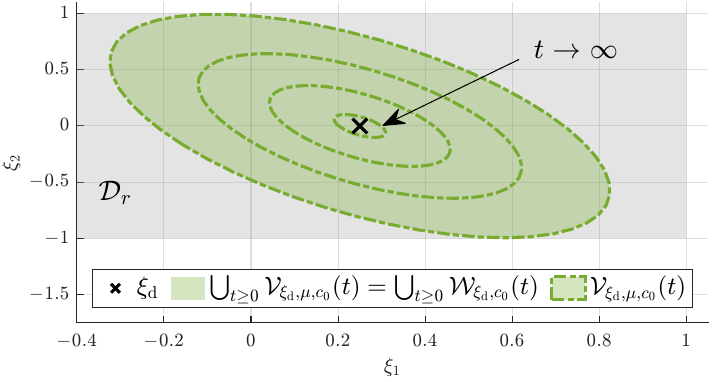}
	\caption{
		Stability analysis for set-point tracking.
		The analysis that incorporates the transient decrease of the tracking error simplifies to the conventional local stability analysis.
	}
	\label{fig:lyapunov_redesign_setpoint}
\end{figure}

Let the set-point $\xi_{\mathrm{d}} = [0.25,0]^\top$.
Figure~\ref{fig:lyapunov_redesign_setpoint} shows the union \eqref{eq:union_in_D} and the set \eqref{eq:invariant_set_extern} for four different times $t \in \{0,1.5,3.5,8\}$s and $c_0 = 0.33$ in the phase-plane. 
Notably, the bound $\mathcal{V}_{\xi_{\mathrm{d}},\mu,c_0}(t) \ni \xi(t)$ decreases over time, thereby illustrating~\eqref{eq:set_point_simplified_union} and~\eqref{eq:set_point_simplified_bounds}.
Moreover, since $\mathcal{V}_{\xi_{\mathrm{d}},\mu,c_0}(0) \subseteq \mathcal{D}_r$, Corollary~\ref{cor:set_point} guarantees tracking for all $\xi_0 \in \mathcal{V}_{\xi_{\mathrm{d}},\mu,c_0}(0)$, where \eqref{eq:bound_solution} characterises the convergence of $\xi$ to the set-point.

\section*{Appendix}
To introduce the comparison lemma, let the continuous function $g:[0,\infty) \times [0,\infty) \to \mathbb{R}$ be such that the scalar differential equation
\begin{align}\label{eq:ode_appendix}
	\dot{\psi} = g(\psi,t),
	\qquad
	\psi(0) \geq 0
\end{align}
admits a unique solution for all $t \geq 0$ for every $\psi(0) \geq 0$.

\begin{lemma}[Comparison Lemma \cite{LakL1969}]\label{lem:comparison_lemma}
	Given the unique solution $\psi$ of \eqref{eq:ode_appendix} and some $T > 0$, let $\chi : [0,T) \to \mathbb{R}$ be a continuously differentiable function
	whose time derivative satisfies $\dot{\chi}(t) \leq g(\chi(t),t)$ for all $t \in [0, T)$ and $\chi(0) \leq \psi(0)$.
	Then, $\chi(t) \leq \psi(t)$ for all $t \in [0,T)$.
\end{lemma}

\bibliographystyle{ieeetr}
\bibliography{literatur.bib}

\end{document}